\documentclass[10pt,twocolumn,english,aps,prl,superscriptaddress]{revtex4}
\usepackage{graphicx,psfrag,amsmath,amssymb,float}
\usepackage{epstopdf}
\usepackage{color}
\input{epsf}
\usepackage{graphicx,natbib,ulem}
\usepackage[colorlinks,bookmarks=false,citecolor=blue,linkcolor=red,urlcolor=blue]{hyperref}

\newcommand{\be}{\begin{equation}}
\newcommand{\ee}{\end{equation}}
\newcommand{\bea}{\begin{eqnarray}}
\newcommand{\eea}{\end{eqnarray}}

\newcommand{\bc}{\begin{center}}
\newcommand{\ec}{\end{center}}
\begin{document}
{{
\title{Impurity-induced magnetic order in low dimensional spin gapped materials}
\author{J. Bobroff}
\affiliation{Laboratoire de Physique des Solides,
Universit\'e Paris-Sud, UMR-8502 CNRS, 91405 Orsay, France}
\author{N. Laflorencie}
\affiliation{Laboratoire de Physique des Solides,
Universit\'e Paris-Sud, UMR-8502 CNRS, 91405 Orsay, France}
\author{L. K. Alexander}
\affiliation{Laboratoire de Physique des Solides,
Universit\'e Paris-Sud, UMR-8502 CNRS, 91405 Orsay, France}
\author{A. V. Mahajan}
\affiliation{Department of Physics, Indian Institute of Technology Bombay, Mumbai 400076, India}
\author{B. Koteswararao}
\affiliation{Department of Physics, Indian Institute of Technology Bombay, Mumbai 400076, India}
\author{P. Mendels}
\affiliation{Laboratoire de Physique des Solides,
Universit\'e Paris-Sud, UMR-8502 CNRS, 91405 Orsay, France}
%
\begin{abstract}
We have studied the effect of non-magnetic Zn impurities in the coupled spin-ladder Bi(Cu$_{1-{\rm x}}$Zn$_{\rm x}$)$_2$PO$_6$ using $^{31}$P NMR, muon spin resonance ($\mu$SR) and Quantum Monte Carlo simulations. Our results show that the impurities induce in their vicinity antiferromagnetic polarizations, extending over a few unit cells. At low temperature, these extended moments freeze in a process which is found universal among various other spin-gapped compounds: isolated ladders, Haldane or Spin-Peierls chains. This allows us to propose a simple common framework to explain the generic low-temperature impurity induced freezings observed in low dimensional spin-gapped materials.
\end{abstract}

\pacs{75.10.Pq, 76.60.-k, 76.75.+i,05.10.Cc}

\maketitle
The physics of low-dimension quantum antiferromagnets (AF) is fascinating and surprising. Simple AF spin-chains or ladders display exotic low temperature behavior such as spin-liquids, spin-gaps, magnetic orders, spin glasses, etc. The precise behavior depends on the value of the spin, the dimensionality of the material, the anisotropy, the relative strengths and signs (possibly frustrating) of the  magnetic couplings. Studying the effect of impurities is an efficient way to classify and reveal the quantum nature of these systems~\cite{Alloul2009}. For example, the
qualitatively different properties of half-integer and integer spin chains are evidenced in the strongly different impurity-induced effects in the two systems.

In AF chains and ladders, non-magnetic impurities induce paramagnetic clouds made of alternating moments in their vicinity, the shape and extension of which is directly linked to the type of electronic correlations involved in the ground state. At low enough temperatures, impurities may even lead to magnetic order or spin freezing, a sort of "order by disorder"
phenomenon~\cite{Kivelson91,Hase93,Azuma97,Uchiyama99,Oosawa2002,Manabe98,Grenier98}.
For example, a few $\%$ of non-magnetic Zn or Mg substituted at the Cu site of the spin-Peierls chain CuGeO$_3$~\cite{Hase93,Manabe98,Grenier98}, the spin-ladder SrCu$_2$O$_3$~\cite{Azuma97,Fujiwara98} or at the Ni site of the Haldane chain PbNi$_2$V$_2$O$_8$~\cite{Uchiyama99, Masuda2002}, leads to a collective freezing  below a 3D ordering temperature T$_{\rm g}$ of a few K. The induced paramagnetic clouds can be viewed as effective localized moments extending over a finite region of typical size $\xi$ around each impurity, with an exponentially decaying staggered magnetization $\langle S^{z}(r)\rangle\sim (-1)^r\exp(-r/\xi)$~\cite{White94,Martins97,Sandvik97}.
A 3D effective interaction between them governs the  freezing at low temperature. The frozen-state characteristics should then be specific to a given compound and geometry, explaining why no common picture has been given yet  to understand the numerous experiments. As the effective interaction between impurities falls off exponentially with the distance $r$~\cite{Sigrist96,Imada97,Yasuda2002,Laflo03-04} as
$
J^{\rm eff}(r)\sim J_0\exp(-r/\xi)$, it has then been suggested that the relevant energy scale for the magnetic freezing is given by the {\it{typical coupling}}, i.e. the effective interaction occurring at the average distance between impurities along the chains or the ladders $\langle r\rangle\simeq 1/x$~\cite{Imada97,Manabe98,Ohsugi99} where $x$ is the impurity concentration. However, as pointed out in Ref.~\onlinecite{Fabrizio99}, realistic parameters would give exponentially small ${\rm{T}}_{\rm g}$,
and a quantitative agreement with experimental values has been obtained using unrealistic enhanced lengths $\xi$~\cite{Manabe98,Ohsugi99}.

To decide which actual process drives the freezing in impurity-doped spin-gaped materials, we have chosen to study the effect of impurities in a new spin ladder {material BiCu$_2$PO$_6$ (BCPO). In contrast with the archetypal ladder compound SrCu$_2$O$_3$ where ladders are {almost isolated with a magnetic coupling $J_{\rm leg}\simeq 2000$ K, BCPO has a non-negligible inter-ladder coupling but a much smaller  $J_{\rm leg}\sim 100$ K~\cite{Koteswararao2007}. We present here $\mu$SR and NMR studies of the effects of non-magnetic Zn impurities, together with Quantum Monte Carlo (QMC) simulations. Above T$_{\rm g}$, NMR enables one to measure the induced paramagnetic cloud near the impurities while $\mu$SR is an ideal probe of the frozen state, measuring both the freezing temperature T$_{\rm g}$ and the corresponding field distribution in the full volume of the sample. Our results compared to QMC simulations demonstrate that the extension of the induced clouds is only a few unit cells. This small extension, {together with the 3D coupling $J_{\rm 3D}$, {are the crucial parameters in the freezing process. We {thus explain {in a common framework the freezing temperatures observed in ladders and other low dimension spin-gapped materials.

Synthesis and structural characterization of Bi(Cu$_{1-{\rm x}}$Zn$_{\rm x}$)$_2$PO$_6$ are presented elsewhere~\cite{Koteswararao2007,Alexander2009}. NMR measurements were performed using a 7 Tesla homemade spectrometer with standard pulse techniques and Fourier Transform recombinations. Magnetic susceptibility was measured in a SQUID magnetometer at H=0.1 T. When Zn is substituted at Cu site, it is expected to release a free $S=1/2$ spin. Macroscopic susceptibility indeed displays the signature of this free spin through a low temperature Curie component proportional to Zn content. The fitted Curie constant corresponds to a total induced spin $S=0.35-0.45$ per Zn slightly smaller than the expected $S=1/2$ because of possible frustration effects ~\cite{Alexander2009}. This total spin is not localized on a single site but develops as an induced, extended, alternating cloud (AF cloud) along the ladder and in adjacent ones. This AF cloud reveals itself in the $^{31}$P NMR spectrum through a low temperature broadening (Fig.~\ref{fig:NMR}) just as in isolated ladders~\cite{Fujiwara98}.
\begin{figure}
\begin{center}
\includegraphics[width=\columnwidth,clip]{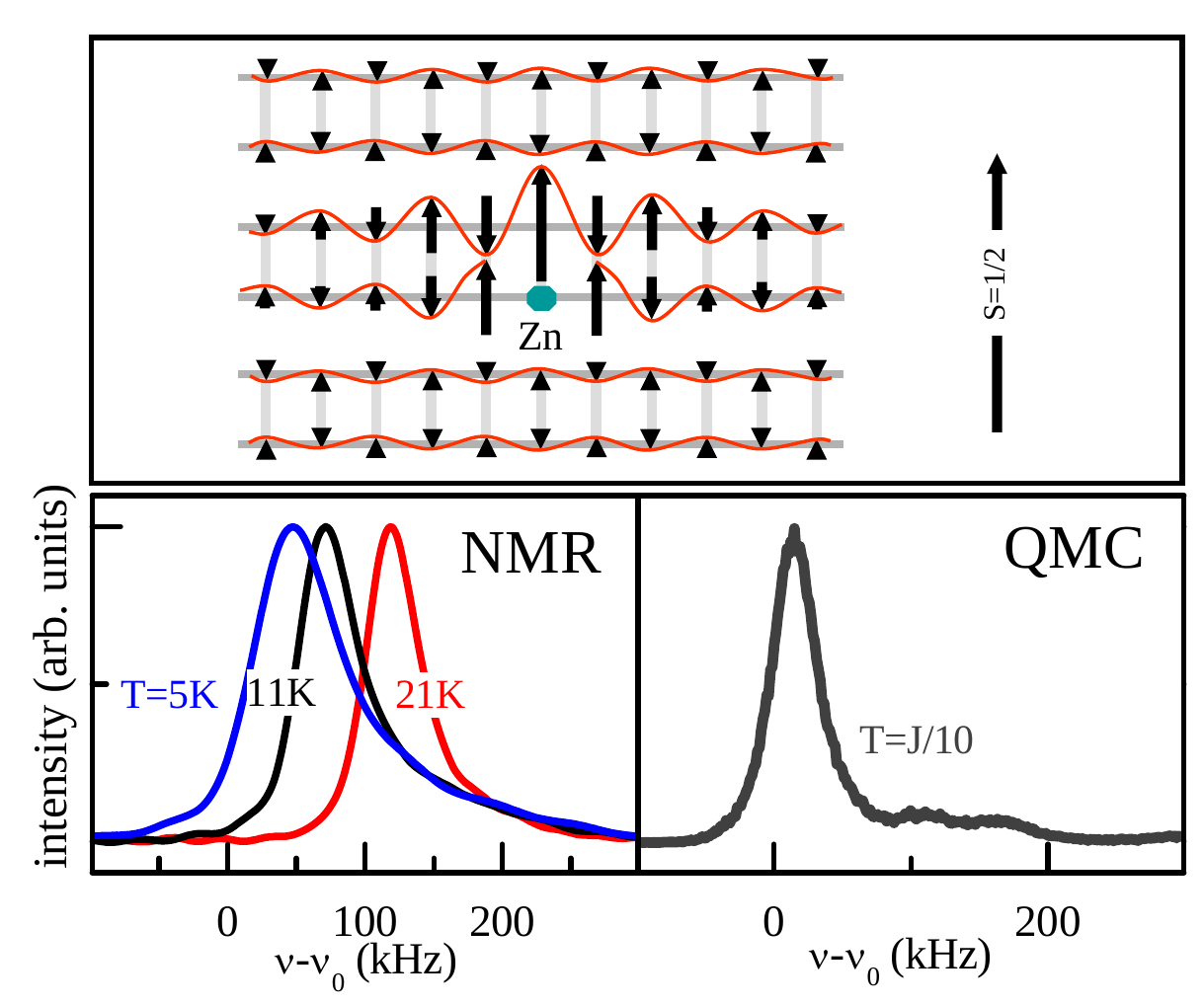}
\end{center}
\caption{(Color online). Lower panel: Left: $^{31}$P NMR spectra for BCPO: Zn 2$\%$ at H=7 T; Right: QMC results for the coupled ladders model Eq.~(\ref{eq:model}) at $T=J/10$ with an applied field $H=J/10$, using $J_{\rm leg}=J_{\rm rung}=J$, $J_{\perp}/J=0.1$, ${\rm x}=2\%$. The corresponding Zn induced magnetic pattern in the ladders is displayed in the top panel, with a S=1/2 arrow indicating the scale.}
\label{fig:NMR}
\end{figure}
{
The spin-$\frac{1}{2}$ coupled ladders can be modeled by:
\bea
{\cal{H}}&=&\sum_{\langle i j\rangle}J_{\rm leg}{\vec{S}}_{i,j}\cdot{\vec{S}}_{i+1,j}+J_{\rm rung}{\vec{S}}_{i,2j}\cdot{\vec{S}}_{i,2j+1}\nonumber \\
&+&J_{\perp}{\vec{S}}_{i,2j+1}\cdot{\vec{S}}_{i,2j+2}.
\label{eq:model}
\eea
As already studied in details~\cite{Matsumoto2001,NewO3}, this model displays a gapped valence bond solid (VBS) phase when the inter-ladder coupling $J_{\perp}$ is not too strong. For isotropic ladders ($J_{\rm rung}=J_{\rm leg}$)
this is the case when
$J_{\perp}<J_{\perp}^{c}$ with $J_{\perp}^{c}/J=0.31407(5)$~\cite{Matsumoto2001}, and the correlation length diverges close to the quantum critical point (QCP) $\xi\sim (J_{\perp}^{\rm c}-J_{\perp})^{-\nu}$ with $\nu= 0.709(6)$~\cite{NewO3}.
Previous bulk studies on BCPO~\cite{Koteswararao2007} estimate $J_{\rm leg}\simeq J_{\rm rung}\sim 100$ K. Muffin tin orbital calculations suggest the existence of an additional frustrating coupling along the ladders~\cite{Koteswararao2007}. We did not take this possible frustration into account in our model Eq.~(\ref{eq:model}) as we just want to capture semi-quantitative features of the low-T physics.
The experimental spin gap $\Delta\sim 35$K yields an inter-ladder coupling $0.1\le J_{\perp}/J_{\rm leg}\le 0.2$, implying a small $\xi$.
\begin{figure}
\begin{center}
\includegraphics[width=\columnwidth,clip]{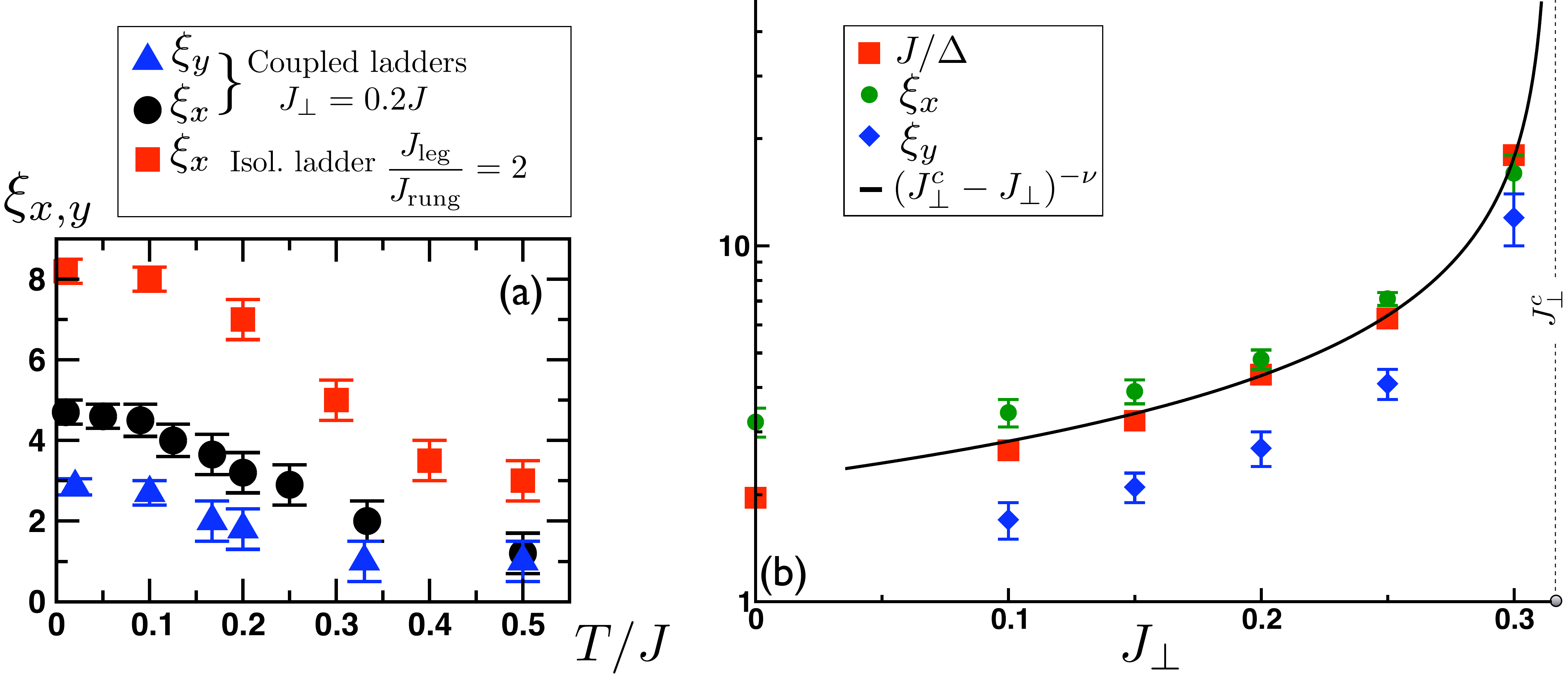}
\end{center}
\caption{{(Color online). QMC results for the AF cloud extensions and the correlation length of model (1). (a) $T$ dependence of the induced extensions $\xi_x$ and $\xi_y$ for coupled and isolated ladders (SrCu$_2$O$_3$ has $J_{\rm leg}/J_{\rm rung}=2$). (b) $T=0$ results for the inverse gap $J/\Delta\sim\xi$ and
the extensions $\xi_x$, $\xi_y$ versus $J_{\perp}$. The black line is the pure correlation length $\xi\sim(J_{\perp}^{c}-J_{\perp})^{-\nu}$ (see text). The small error bars come from the uncertainty in the 3 parameter fit of QMC data to Eq.(\ref{eq:profile}).}}
\label{fig:corr}
\end{figure}
Using QMC simulations we have computed the magnetization profile induced by a single impurity on a $32\times 32$ lattice.
Exponentially localized AF profiles are found, with the 2D form
\be
\langle S^{z}({\vec{r}})\rangle\simeq S_0\exp({-\frac{x}{\xi_x}-\frac{y}{\xi_y}})(-1)^{x+y}
\label{eq:profile}
\ee
valid all over the gapped regime. The extensions $\xi_x$ and $\xi_y$ at $T=0$ are reported in Fig.~\ref{fig:corr}(b) where one can clearly see that they quantitatively follow the behavior of the pure correlation length $\xi\sim J/\Delta$:
the impurity induced effect reveals the intrinsic properties of the ladder, as in Haldane chains~\cite{Tedoldi99,Das2004} or high T$_{\rm c}$ cuprates~\cite{Ouazi2004}. This result contradicts analysis of Ref.~\onlinecite{Ohsugi99} which argues that the AF cloud extension depends on the impurity content and reaches values as large as 50 cell units at low concentrations. However, such conclusions were entirely based on NMR experiments done at the very low x$=0.1\%$ concentration where the observed NMR broadenings are too small to be reliable. All other data of Refs.~\onlinecite{Ohsugi99} and \onlinecite{Fujiwara98} are fully compatible with our own results and conclusions. {As shown in Fig.~\ref{fig:corr}(a), the cloud extensions $\xi_x$ and $\xi_y$ increase when the temperature is lowered but converge very rapidly once the gap $\Delta\simeq 0.2-0.3J$ is reached. As a result, the NMR broadening (Fig.~\ref{fig:NMR}) at low T is expected to follow the Curie law due to $S_{0}$ only. This is indeed the case in the experiments presented in Fig.~\ref{fig:NMR}. QMC simulations performed for $J_{\perp}/J=0.1$ at $T=J/10$ with an external field $H=J/10$ over $32\times 32$ sites with x$=2\%$ of non-magnetic impurities are averaged over $10^2$ disordered samples to obtain the distribution of local fields seen by each phosphorus plotted in the right panel of Fig.~\ref{fig:NMR}. The simulated broadening found without any adjustable parameter is remarkably similar to the experimental ones, albeit possible frustration is not considered.


\begin{figure}
\begin{center}
\includegraphics[width=\columnwidth,clip]{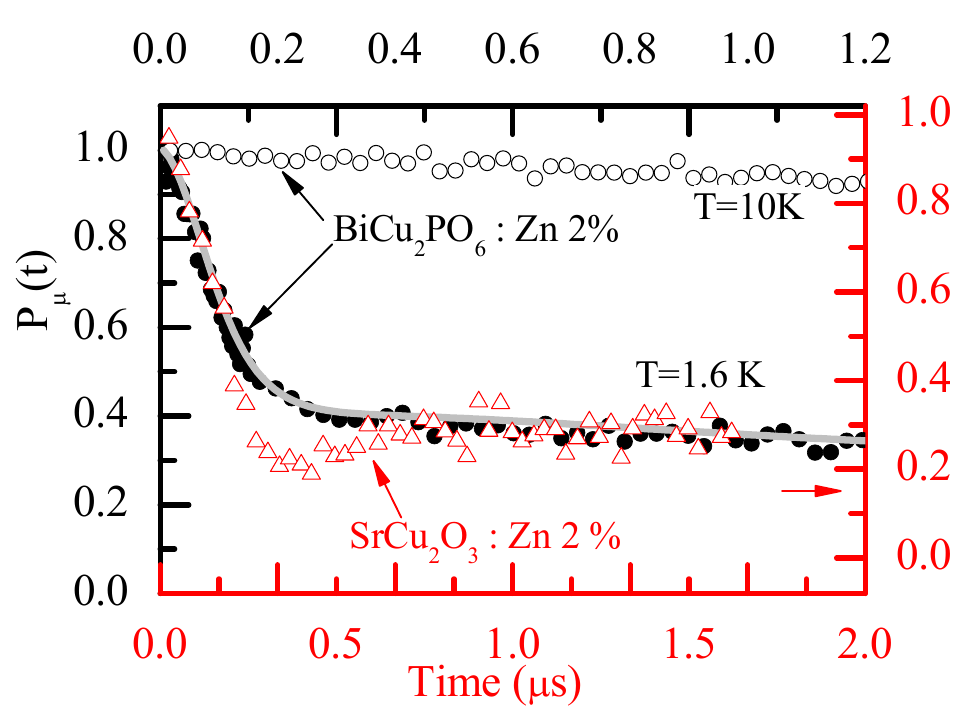}
\end{center}
\caption{(Color online). Muon polarization for BCPO above and below T$_{\rm g}$  (black circles) on upper and left axis is compared to a similar experiment made in SrCu$_2$O$_3$ (bottom axis being rescaled, right axis is slightly shifted because of a different background) from Ref.~\onlinecite{Larkin2000}.
Grey line is a fit to Eq.3 of Ref.~\onlinecite{Larkin2000} as explained in the text.}
\label{fig:muSR}
\end{figure}

In order to probe whether these impurity induced AF clouds of only a few unit cell extension lead to a freezing in BCPO, we performed a $\mu$SR study on the same Zn substituted BCPO samples at the PSI (GPS) facility. Pressed disks of randomly oriented powders were used. Muons which carry a spin 1/2 are implanted in the samples and precess around the local magnetic field. This precession is measured through the detection of the asymmetry of the positron emission due to the muon decay. In the absence of an external magnetic field ("zero field setup"), above T$_{\rm g}$, only the tiny nuclear spin dipolar fields are experienced by the muons. This results in a gaussian field distribution and a very slowly-decaying "Kubo-Toyabe" polarization as displayed in Fig.~\ref{fig:muSR} at T=10K. When the electronic spins start to freeze below T$_{\rm g}$, their randomly oriented static moments result in a much larger field distribution and a much faster decaying asymmetry as shown in Fig.~\ref{fig:muSR} at T=1.6 K. We checked that this depolarization is of static origin below T$_{\rm g}$  by applying a large longitudinal field and finding the expected asymmetry decoupling. Since no marked oscillations are observed, the corresponding field distribution is not that of a simple commensurate magnetic ordering. However, due to the lack of a dip in the asymmetry, it is probably not completely random as in a spin glass, but more likely in an intermediate situation with AF clusters, as discussed in Ref.~\onlinecite{Larkin2000}. To get an estimate of the field distribution, we used the same  phenomenological exponential field distribution as in Ref.~\onlinecite{Larkin2000}, which fits well the data as shown in Fig.~\ref{fig:muSR}. The resulting static field distribution develops as a mere order parameter which allows to determine T$_{\rm g}$. Whatever the impurity nature or concentration, the transition temperature is proportional to the field distribution even for very large impurity concentrations.

The SrCu$_2$O$_3$ ladder shows similar time dependence for the  muon polarization as displayed in Fig.~\ref{fig:muSR}~\cite{Larkin2000}. Both the static field distribution and T$_{\rm g}$~\cite{Azuma97} are found 1.7 times larger than in BCPO. This 1.7 ratio is unexpectedly about one order of magnitude smaller than the ratio between the ladder couplings. Other low-D spin gap systems such as Haldane chains or Spin-Peierls chains also display very similar T$_{\rm g}$  despite their different geometries and gap values (Fig.~\ref{fig:Tg}).
The dependence on impurity content also shows a strikingly generic behavior, with a similar departure from a linear behaviour at about 2-3$\%$ for all compounds as demonstrated in the right panel of Fig.~\ref{fig:Tg}. Magnetic impurities in these materials also lead to similar T$_{\rm g}$  and similar concentration dependence as shown in Fig.~\ref{fig:Tg}.

\begin{figure}
\begin{center}
\includegraphics[width=\columnwidth,clip]{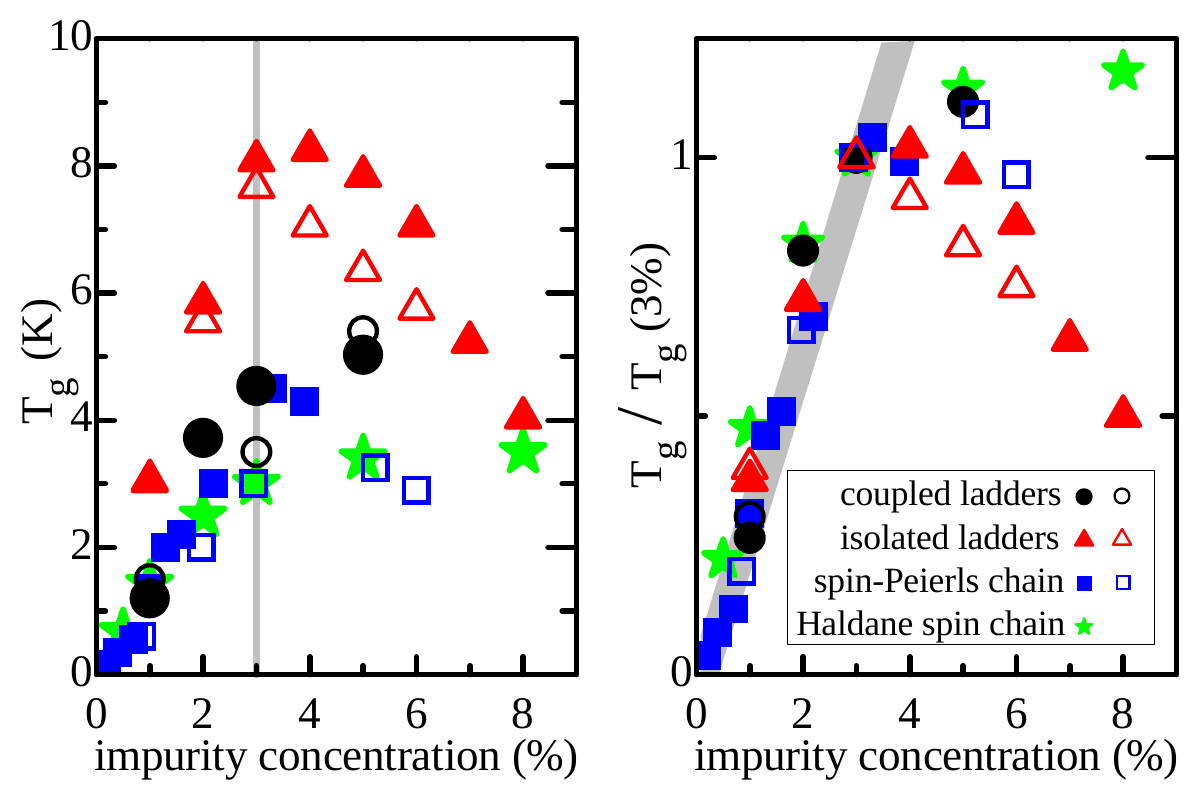}
\end{center}
\caption{(Color online) Left panel: transition temperatures versus impurity concentration for various low-D spin-gapped systems: coupled ladders Bi(Cu$_{1-x}$(Zn or Ni)$_x$)$_2$PO$_6$ from this study; isolated ladder Sr(Cu$_{1-x}$(Zn or Ni)$_x$)$_2$O$_3$ ~\cite{Azuma97}~\cite{Ohsugi99}; Haldane chain Pb(Ni$_{1-x}$Mg$_x$)$_2$V$_2$O$_8$~\cite{Imai2004}; spin-Peierls chains Cu$_{1-x}$(Zn or Ni)$_x$GeO$_3$ ~\cite{Grenier98}. Full and open symbols correspond respectively to non magnetic and magnetic impurities. Right panel: Same data where T$_{\rm g}$  is rescaled by its value at x=3$\%$.}
\label{fig:Tg}
\end{figure}

{Despite several theoretical investigations devoted to understand the origin of the impurity-induced 3D ordering in low-D gapped systems~\cite{Imada97, Troyer97,Greven98,Fabrizio99,Laflo03-04}, no common framework has emerged so far which explains this generic T$_{\rm g}$ behavior. The collective freezing of the effective moments (having a 3D extension $\sim\xi_x\xi_y\xi_z$ at $T>T_g$) is actually controlled by the exponentially decaying 3D coupling of the general form
\be
|J_{\rm 3D}^{\rm eff}({\vec{r}})|\simeq J_{\rm 3D}\exp\left(-\frac{x}{\xi_x}-\frac{y}{\xi_y}-\frac{z}{\xi_z}\right),
\label{eq:eff}
\ee
expected to occur for the wide class of spin gapped materials~\cite{Sigrist96,Imada97,Yasuda2002,Laflo03-04,Wessel2001,Vojta2006}.
Couplings in the three directions are necessary to allow finite-T ordering due to the Mermin-Wagner theorem, and $J_{\rm 3D}$ is the weakest of these couplings. The typical coupling $J_{\rm typ}=|J_{\rm 3D}^{\rm eff}(\langle r\rangle)|$, taken at the average distance $\langle r\rangle$ between impurities, is much too small to explain the actual 3D freezing temperatures ~\cite{Fabrizio99} because it does not take into account the rare but crucial situations where $r$ is small. On the contrary, the average coupling $J_{\rm av}$ taken over all possible $J_{\rm 3D}^{\rm eff}({\vec{r}})$ does account for the broad distribution of effective interactions and is just given by
\be
J_{\rm av}=\langle |J_{\rm 3D}^{\rm eff}({\vec{r}})|\rangle \simeq J_{\rm 3D}\frac{{\rm x}V_{\xi}}{1+{\rm x}V_{\xi}}
\label{eq:Javg}
\ee
where $V_{\xi}\sim\xi_x\xi_y\xi_z$ is the magnetic volume occupied by each induced moment. We propose that this average coupling governs the ordering, i.e.  ${\rm{T}}_{\rm g}\simeq J_{\rm av}$. Indeed, this model accounts well for the fact that all ${\rm{T}}_{\rm g}$ are similar in magnitude, because both $J_{\rm 3D}$ and $V_{\xi}$ are roughly of the same order of magnitude among the different systems. Precise values for $J_{\rm 3D}$ are not known since they are at most a few percent of the dominant coupling. However, rough estimates give a few K for CuGeO$_3$~\cite{Regnault96} and PbNi$_2$V$_2$O$_8$~\cite{Uchiyama99}. For SrCu$_2$O$_3$ a theoretical calculation~\cite{Troyer97} based on an isotropic ladder model found $J_{\rm 3D}\sim 20 K$. Even though the order of magnitude is correct, the actual $J_{\rm 3D}$ is probably smaller because of the anisotropy ${J_{\rm leg}}/{J_{\rm rung}}=2$ ~\cite{Johnston96}. Regarding BCPO, based on our analysis we can confidently predict a 3D coupling $\sim 5K$.
Turning now to the magnetic volumes $V_{\xi}$ of the induced moments, they are similar at low temperature, typically $\sim 30$ sites, despite the geometric differences between systems. For instance, Zn in the almost isolated anisotropic ladders Sr(Cu$_{1-x}$Zn$_x$)$_2$O$_3$ induces a cigar-like pattern, with  $V_{\xi}\simeq 2\times 8\times 2\simeq 32$. In the coupled ladders Bi(Cu$_{1-x}$Zn$_x$)$_2$PO$_6$ with $J_{\perp}/J\simeq 0.15$, Zn induces a pancake-like volume $V_{\xi}\simeq2\times 3.9\times 2\times 2.1\simeq 33$. This is because all these spin-gapped materials are far enough from a QCP so that the various $\xi$ remain a few unit cells and do not diverge (Fig.~\ref{fig:corr}), whereas a totally different physics would be observed at criticality~\cite{QCP}.

Such an analysis allows to understand the concentration dependence of ${\rm{T}}_{\rm g}$ as well. At small x, the average volume between impurities $\langle r\rangle^3 \sim {\rm x}^{-1}$ is large compared to the magnetic volume $V_{\xi}$, and Eq.~(\ref{eq:Javg}) implies a linear behavior of ${\rm{T}}_{\rm g}\simeq J_{\rm av}$ with x. At larger x, extended moments get close enough to percolate and the critical temperature is not linear with x anymore. This should occur at ${\rm x}\sim1/V_{\xi}\simeq 1/30\simeq 3\%$. This prediction of a linear increase of ${\rm{T}}_{\rm g}$ below ${\rm x}\simeq 3\%$ and a deviation above is exactly what is observed in Fig.~\ref{fig:Tg} (see also the attached supplementary material where QMC simulations are presented).

In summary, the framework we propose to explain the impurity induced freezings in various systems is based on the fact that the impurity induces an extended magnetic moment, the volume of which is similar in all systems because all lie far from the QCP. These induced moments interact through a 3D interaction which is not much sensitive to the geometry or the gap nature. This holds for magnetic impurities as well. It will be interesting to understand if this universal picture survives when a Quantum Critical Point is approached.

We acknowledge enlightening discussions with F. Alet, H. Alloul, M. Azuma, S. Capponi, I. Dasgupta, B. Grenier, Y. Kitaoka, T. Masuda, V. Simonet, A. Zorko and we thank A. Amato for technical support at PSI facility. This work was supported by the EC FP 6 programme, Contract No. RII3-CT-2003-505925, ANR Grant No. NT05-4-41913 "OxyFonda", the Indo-French center for the Promotion of Advanced Research and ARCUS Ile de France. NL acknowledges LPT (Toulouse) for hospitality.

\end{document}